\documentclass[a4paper,fleqn,usenatbib]{mnras}
\usepackage{amsmath}
\usepackage{epsfig} 
\usepackage{graphicx}	
\usepackage{amssymb}	
\newcommand{\be}{\begin{equation}}
\newcommand{\e}{\end{equation}}
\newcommand{\bear}{\begin{eqnarray}}
\newcommand{\ear}{\end{eqnarray}}

\newcommand{\hmpc}{{\, h^{-1}\, {\rm Mpc}}}

\def\aj{AJ}
\def\apj{ApJ}
\def\apjs{ApJS}

\def\mnras{MNRAS}
\def\aap{A\&A}

\def\apjs{ApJS}
\def\apjl{ApJ Letters}
\def\physrep {Physics Reports}

\title[Relating entropy and mass variance] {Relating information
  entropy and mass variance to measure bias and non-Gaussianity}
    
\author[Pandey, B.]  { Biswajit Pandey\thanks{E-mail:
    biswap@visva-bharati.ac.in} \\ Department of Physics,
  Visva-Bharati University, Santiniketan, Birbhum, 731235, India\\ }
   
 \date{\today}

 \pubyear{2016}
  
\begin{document}
\label{firstpage}
\pagerange{\pageref{firstpage}--\pageref{lastpage}}      
\maketitle
       
\begin{abstract}

We relate the information entropy and the mass variance of any
distribution in the regime of small fluctuations. We use a set of
Monte Carlo simulations of different homogeneous and inhomogeneous
distributions to verify the relation and also test it in a set of
cosmological N-body simulations. We find that the relation is in
excellent agreement with the simulations and is independent of number
density and the nature of the distributions. We show that the relation
between information entropy and mass variance can be used to determine
the linear bias on large scales and detect the signatures of
non-Gaussianity on small scales in galaxy distributions.

\end{abstract}


\section{Introduction}

Understanding the formation and evolution of the large scale
structures in the Universe is one of the most complex issues in
cosmology. The galaxies are the basic building blocks of the large
scale structures and their spatial distribution reveals how the
luminous matter is distributed in the Universe. The study of the
distribution of galaxies is one of the most direct probe of the large
scale structures. The modern galaxy redshift surveys like the Sloan
Digital Sky Survey (SDSS) \citep{york} has now mapped the distribution
of more than a million galaxies and quasars which provides the most
detailed three-dimensional maps of the Universe ever made in the
history of mankind. The maps reveal that the galaxies are distributed
in an interconnected complex filamentary network namely the cosmic
web. The cosmic web emerges naturally from the gravitational
amplifications of the primordial density fluctuations seeded in the
early universe. The distribution of galaxies in the cosmic web encodes
a wealth of information about the formation and evolution of the large
scale structures. A large number of statistical tools have been
developed so far to quantify the galaxy distribution and unravel the
large scale structures. The correlation functions \citep{peeb80}
characterize the statistical properties of the galaxy distributions.
The two-point correlation function and its Fourier space counterpart,
the power spectrum remain some of the most popular measure of galaxy
clustering todate. These statistics provide a complete description for
the primordial density perturbations which are assumed to be Gaussian
in the linear regime. But in subsequent stages of non-linear
gravitational evolution, the phase coupling of the Fourier modes
produces non-vanishing higher-order correlation functions and
polyspectra. In principle a full hierarchy of N-point statistics is
required to provide a complete description of the distribution. The
void probability function \citep{sdm} provides a characterization of
voidness that combine many higher moments of the distribution. Other
methods to quantify the cosmic web includes the percolation analysis
\citep{shand1,einas1}, the genus statistics \citep{gott}, the minimal
spanning tree \citep{barrow}), the Voronoi tessellation
\citep{ike,weygaert}, the Minkowski functionals \citep{mecke,smal},
the Shapefinders \citep{sahni}, the critical point statistics
\citep{colombi}, the marked point process \citep{stoi1}, the
multiscale morphology filter \citep{arag}, the skeleton formalism
\citep{sous} and the local dimension \citep{sarkar}.

The popularity of the two-point correlation function and the power
spectrum lies in the fact that they can be easily measured and related
to the theories of structure formation whereas it is hard to do so for
most of the other statistics. The variance of the mass distribution
smoothed with a sphere of radius $r$ is a simple and powerful
statistical measure which is directly related to the power spectrum.
Information entropy is a statistical measure which can help us to
study the formation and evolution of structures from an information
theoretic viewpoint. Recently information entropy has been used as a
measure of homogeneity \citep{pandey13,pandey15,pandey16a} and
isotropy \citep{pandey16b} of the Universe. Both the information
entropy and the mass variance can be used as a measure of the
non-uniformity of a probability distribution. The entropy uses more
information about the probability distribution as it is related to the
higher order moments of a distribution. The variance can be treated as
an equivalent measure only when the probability distribution is fully
described by the first two moments such as in a Gaussian
distribution. However a highly tailed distribution is not uniquely
determined even by its all the higher order moments
\citep{patel,romano,carron2,carron1}.

Different statistical tools have been designed to explore different
aspects of the galaxy distribution and when possible it is important
to relate these statistical measures for a better interpretation of
different cosmological observations. In future, Information theory can
find many potential applications in cosmology. It would be highly
desirable to relate the information entropy to the other conventional
measures such as the mass variance and the power spectrum. In this
paper we particularly explore the relation between the discrete
Shannon entropy and the mass variance of a distribution and test it
using N-body simulations and the Monte Carlo simulations of different
distributions.

We also explore a few possible applications of this relation in
cosmology particularly to constrain the large-scale linear bias and
non-Gaussianity in galaxy redshift surveys. Galaxies are known to be a
biased tracer of the underlying dark matter distribution. Currently
there exist several methods to determine the large scale linear
bias. The bias can be directly determined from the two-point
correlation function and power spectrum \citep{nor,teg,zehavi10}, the
three-point correlation function and bispectrum \citep{feldman,
  verde,gaztanaga}, the redshift space distortion parameter
$\beta=\Omega_m^{0.6}/b$ \citep{haw,teg} and the filamentarity of the
galaxy distribution \citep{bharad,pandeyb}. We employ the information
entropy-mass variance relation to propose a new method which can be
used to determine the large scale linear bias from galaxy redshift
surveys.

Non-Gaussianity of the cosmic density field is one of the most
interesting issues in cosmology. In the current paradigm, the
primordial density fluctuations are assumed to be Gaussian and
signatures of non-Gaussianities in these fluctuations can be used to
constrain different inflationary models in cosmology
\citep{bartolo}. As the density fluctuations grow, the probability
distribution function of the cosmic density field develope extended
tails in the overdense regions and gets truncated in the underdense
regions. These non-Gaussianities induced by the structure formation
are much stronger and dominates any primordial non-Gaussianities from
the early Universe. In the present work we do not address the
primordial non-Gaussianities but the non-Gaussianties introduced by
the nonlinear evolution of the cosmic density field. We investigate if
the information entropy-mass variance relation can be used to detect
the signatures of non-Gaussianity in the galaxy distribution.

A brief outline of the paper follows. In section 2 we describe the
relation between information entropy and mass variance followed by a
discussion on some possible applications of this relation in section
3. We describe the data in section 4 and finally present the results
and conclusions in section 5.


\section{Information entropy and cosmological mass variance}

Information entropy \citep{shannon48} is a measure in Information
theory which quantify the amount of information required to describe a
random variable. According to Shannon, the information contained in
any outcome of a probabilistic process is given by
$I(p(x_i))=\log\frac{1}{p(x_i)}$ where $p(x_i)$ is the probability of
that particular outcome. If there are $n$ outcomes of the random
variable $x$ given by $\{x_{i}:i=1,....n\}$ and $N$ observations are
made then we expect $N\,p(x_i)$ occurrences for each of the outcome
$x_i$. The average information required to describe the discrete
random variable $x$ is then given by the information entropy $H(x)$
defined as,

\begin{eqnarray}
H(x) & = & \frac{1}{N}\sum^{n}_{i=1} N\, p(x_i)\,\log \frac{1}{p(x_i)}\\
\nonumber & = & - \sum^{n}_{i=1} \, p(x_i) \, \log \, p(x_i)
\label{eq:shannon1}
\end{eqnarray}

We consider a three dimensional distribution of $N$ points in a finite
region of space. We place a measuring sphere of radius $r$ centered at
each point in the distribution and count the number of other points
$n_i(<r)$ within the sphere. Only the centers for which the measuring
spheres lie completely within the spatial boundary of the region are
considered. To ensure this, we discard all the centers which reside
within a distance $r$ from the boundary. Evidently the number of
centers $M(r)$ available at radius $r$, would decrease with increasing
$r$ for any finite volume. We consider the subset of all the points
which are residing in the $M(r)$ spheres available at any particular
radius $r$. If a point is randomly drawn out of this subset the random
point is expected to reside in one or multiple of these spheres with
different probabilities each given by,
$f_{i,r}=\frac{\rho_{i,r}}{\sum^{M(r)}_{i=1} \, \rho_{i,r}}$ with the
constraint $\sum^{M(r)}_{i=1} \, f_{i,r}=1$. Here
$\rho_{i,r}=\frac{n_{i}(<r)}{\frac{4}{3}\pi r^{3}}$ is the density at
the $i^{th}$ center. We consider this experiment at each radius $r$
and label the outcome with a random variable $x_{r}$ for radius $r$.

The random variable $x_{r}$ has the information entropy,
\begin{eqnarray}
H_{r}& = &- \sum^{M(r)}_{i=1} \, f_{i,r}\, \log\, f_{i,r} \nonumber\\ &=& 
\log(\sum^{M(r)}_{i=1}n_i(<r)) - \frac {\sum^{M(r)}_{i=1} \,
  n_i(<r) \, \log(n_i(<r))}{\sum^{M(r)}_{i=1} \, n_i(<r)}
\label{eq:shannon2}
\end{eqnarray}
where the base of the logarithm is $10$. It may be noted here that the
choice of the base is arbitrary and different choices would only
result in different units for entropy.

The values of $f_{i,r}$ become $\frac{1}{M(r)}$ for all the centres
when the randomly drawn point has the same probability to appear in
any of the $M(r)$ spheres. This maximizes the entropy of $x_{r}$ to
$(H_{r})_{max}=\log \, M(r)$. The maximum entropy corresponds to
maximum uncertainty in the location of the randomly drawn point. The
quantity $(H_{r})_{max}-H_{r}$ at any $r$ quantifies the deviation of
the entropy from its maximum value. We are interested to relate this
statistical measure to other conventional measures of large scale
structures. We calculate $(H_{r})_{max}-H_{r}$ for a distribution with
small fluctuations in the number density. We write the number counts
in \autoref{eq:shannon2} as $n_{i}(<r)=n_{0}(<r)+\delta n_{i}(<r)$,
where $n_{0}(<r)$ is the mean number counts in spheres of radius $r$
and $\delta n_{i}(<r)$ is the fluctuations around the mean. We expand
the $\log(1+\frac{\delta n_{i}(<r)}{n_{0}(<r)})$ terms in Taylor
series as $\frac{\delta n_{i}(<r)}{n_{0}(<r)}<1$. Keeping terms upto
third order and simplifying the expression we get,
\begin{eqnarray}
(H_{r})_{max}-H_{r} \,=\,  \frac{1}{2\,M(r)\,n_{0}(<r)^{2}} \sum^{M(r)}_{i=1}\delta n_i^{2}(<r)  \nonumber \\ - \frac{1}{6\,M(r)\,n_{0}(<r)^{3}} \sum^{M(r)}_{i=1}\delta n_i^{3}(<r) \nonumber \\ +  \frac{1}{3\,M(r)\,n_{0}(<r)^{4}} \sum^{M(r)}_{i=1}\delta n_i^{4}(<r)-...
\label{eq:shannon3}
\end{eqnarray}

The variance is a conventional measure of the non-uniformity present in a
distribution and it is widely used in cosmology. For example one can
estimate the mass variance of the smoothed density field from the
power spectrum of the density field.
\begin{eqnarray} 
\sigma_{r}^{2}\,=\,
  \frac{1}{(2\,\pi)^{2}}\, \int_{0}^{\infty} k^{2} P(k) {W(kr)}^{2} dk
    \label{eq:variance1}
\end{eqnarray}
where, $r$ is the size of the spherical top hat filter, $P(k)$ is the
power spectrum and $W(kr)=3 \frac{\sin(kr)-kr\cos(kr)}{(kr)^{3}}$ is
the Fourier transform of the top-hat window function.

Alternatively one can also estimate the normalized mass variance of
the 3D distribution by simply using the number counts and assuming
equal mass for all the particles.  The normalized mass variance in
this case is given by,\\
\begin{equation}
\sigma_{r}^{2} = \frac{\overline{n^2(<r)}-(\bar{n}(<r))^2}{(\bar{n}(<r))^2}
\label{eq:variance2}
\end{equation}

Here the bars denote average of the respective quantities over the
$M(r)$ spheres available at radius $r$.

Rewriting the number counts in \autoref{eq:variance2} as
$n_{i}(<r)=n_{0}(<r)+\delta n_{i}(<r)$ and simplifying we get,
\begin{eqnarray}
\sigma_{r}^{2}\,=\, \frac{1}{\,M(r)\,n_{0}(<r)^{2}}
\sum^{M(r)}_{i=1}\delta n_i^{2}(<r)
\label{eq:variance3}
\end{eqnarray}

We see from \autoref{eq:shannon3} and \autoref{eq:variance3} that
there is no direct relationship between the entropy and the mass
variance. However in the limit of small fluctuations i.e. when
$\frac{\delta n_{i}(<r)}{n_{0}(<r)}<<1$, one can drop the higher order
terms in \autoref{eq:shannon3} to relate entropy with variance as,
\begin{eqnarray} 
(H_{r})_{max}-H_{r} \,=\, \frac{\sigma_{r}^{2}}{2}
\label{eq:shannvar}
\end{eqnarray}

It may be noted here that in the numerical calculations of entropy
using \autoref{eq:shannon2} we chose the base of the logarithm to be
$10$ whereas the analytical expression obtained in
\autoref{eq:shannon3} is based on the natural logarithm. So a factor
of $\frac{1}{\log_{e} 10}$ should be multiplied to the right hand side
of \autoref{eq:shannon3} and \autoref{eq:shannvar} while comparing
them with the numerical results from simulations.

The maximum entropy $(H_{r})_{max}$ at scale $r$ is not the same for
different distributions as $M(r)$ varies differently with $r$ for
different distributions. In the present scheme $M(r)$ can be written
as,
\begin{eqnarray}
 M(r) =  n_{0} \, \int_{0}^{R-r} (1+\xi(y)) \, d^3y \nonumber\\ =  
 n_{0} \, V + 4 \pi n_{0} \int_{0}^{R-r} y^{2} \, \xi(y) \, dy
\label{eq:clustmean}
\end{eqnarray}
which is just the average number of points in a sphere of radius
$R-r$. Here $R$ is the radius of the entire spherical sampling volume,
$n_{0}$ is the mean density of the distribution, $\xi(y)$ is the two
point correlation function and $V=\frac{4}{3} \pi (R-r)^{3}$ is the
volume of the sphere hosting all the centers of the spheres having a
radius $r$. Behaviour of $\xi(y)$ would be different for different
distributions. Further \autoref{eq:clustmean} would underestimate
$M(r)$ if the higher order correlation functions are non-zero.

We numerically compute the information entropy for different
distributions using \autoref{eq:shannon2} and the mass variance for
the same distributions using \autoref{eq:variance2} and compare them
to test the validity of the relation given by \autoref{eq:shannvar}.

\section{Bias and non-Gaussianity from the information entropy-mass variance relation}

Using the information entropy-mass variance relation described in the
previous section we propose a method for determining the large-scale
linear bias of galaxy distributions from galaxy surveys. On large
scales one can define the linear bias of galaxies by the ratio
$b=\frac{\delta_{g}}{\delta_{m}}$, where $\delta_{g}$ and $\delta_{m}$
are the smoothed density contrast of galaxies and dark matter
respectively. The power spectrum $P(k)$ is given by, $\langle
\delta(\vec{k})
\delta(\vec{k^{\prime}})\rangle=(2\pi)^{3}\delta_{D}^{3}(\vec{k}-\vec{k^{\prime}})\,P(k)$. So
the power spectrum $P_{g}(k)$ of the galaxies would be $b^{2}$ times
the power spectrum $P_{m}(k)$ of the dark matter on large scales.
According to \autoref{eq:variance1}, the mass variance of the galaxy
distribution is also expected to be $b^{2}$ times the mass variance of
the dark matter distribution on large scales. This immediately
suggests that one can use the information entropy-mass variance
relation (\autoref{eq:shannvar}) to determine the linear bias
parameter on large scales. One needs simply the ratios of
$(H_{r})_{max}-H_{r}$ for the galaxy distribution and dark matter
distribution to determine the linear bias given by,
\begin{eqnarray}
b=\sqrt{ \frac {((H_{r})_{max}-H_{r})_{g}}{((H_{r})_{max}-H_{r})_{m}}}
\label{eq:biasval}
\end{eqnarray}
The \autoref{eq:biasval} involves only the measurement of the
information entropy from the galaxy distributions and the N-body
simulations of the $\Lambda$CDM model. It may be noted here that in
this method one can completely bypass the measurements of the power
spectrum $P(k)$, the two-point correlation function $\xi(r)$ and the
mass variance $\sigma_{r}^{2}$ of the associated distributions. The
information entropy is relatively straightforward to compute compared
to the power spectrum or the two-point correlation function. This
provides a simple alternative method to determine the large-scale
linear bias parameter from the measurements of information entropy
alone.

Further the relation can be also used to study the evolution of
clustering in the galaxy distribution by comparing the
$(H_{r})_{max}-H_{r}$ measures at different redshifts. The shape of
the power spectrum is preserved on large scales and its amplitude
grows proportional to $D^{2}(t)$ and $b^{2}(t)$ where $D(t)$ is the
growing mode of density fluctuations and $b(t)$ is the time dependent
bias parameter. Measuring $(H_{r})_{max}-H_{r}$ at different redshifts
may also allow us to constrain $D(t)$ or $b(t)$ provided one of them
is known from other measurements.

One can also detect the signatures of non-Gaussianity in the galaxy
distribution from the measurements of information entropy and
variance. The Gaussian probability distribution function (PDF) is
perfectly symmetrical and it is well known that for a Gaussian PDF all
the odd moments are zero and the $n^{th}$ even moments can be written
in terms of $\sigma^{2n}$ where $\sigma$ is the variance of the
distribution. The \autoref{eq:shannon3} suggests that a condition such
as $(H_{r})_{max}-H_{r}-\frac{\sigma_{r}^{2}}{2}<0$ can occur only in
a non-Gaussian distribution. It may be noted that all the terms with a
negative sign in the expression for
$(H_{r})_{max}-H_{r}-\frac{\sigma_{r}^{2}}{2}$ contain only the odd
moments. So this measure can be negative only if the odd moments are
non-zero and contribution from all the odd moments dominates that from
all the even moments. However the converse is not true because
vanishing of all the odd moments of a probability distribution
function does not always ensure that the probability distribution
function is even \citep{romano}. So a condition such as
$(H_{r})_{max}-H_{r}-\frac{\sigma_{r}^{2}}{2}>0$ does not necessarily
imply that the distribution is not non-Gaussian. The quantity
$(H_{r})_{max}-H_{r}-\frac{\sigma_{r}^{2}}{2}$ thus provide us some
important information on the signatures of non-Gaussianity in the
galaxy distribution.

\begin{figure}
\resizebox{9cm}{!}{\rotatebox{0}{\includegraphics{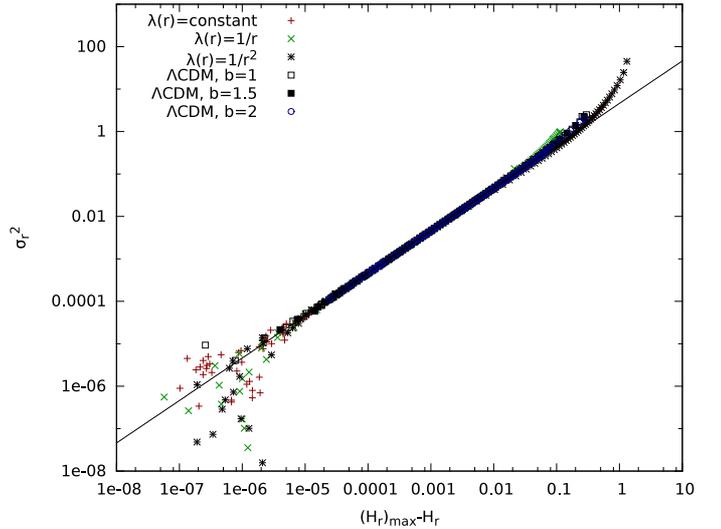}}}\\
\caption{This shows $\sigma_{r}^{2}$ as a function of
  $(H_{r})_{max}-H_{r}$ for the different distributions considered
  here. The solid straight line corresponds to the relation
  $(H_{r})_{max}-H_{r}=\frac{\sigma_{r}^{2}}{4.6}$. Here the factor
  $\frac{1}{4.6}$ comes from multiplying the factor $\frac{1}{2}$ in
  \autoref{eq:shannvar} with $\frac{1}{\log_{e} 10}$.}
  \label{fig:result1}
\end{figure}

\begin{figure*}
\resizebox{7cm}{!}{\rotatebox{0}{\includegraphics{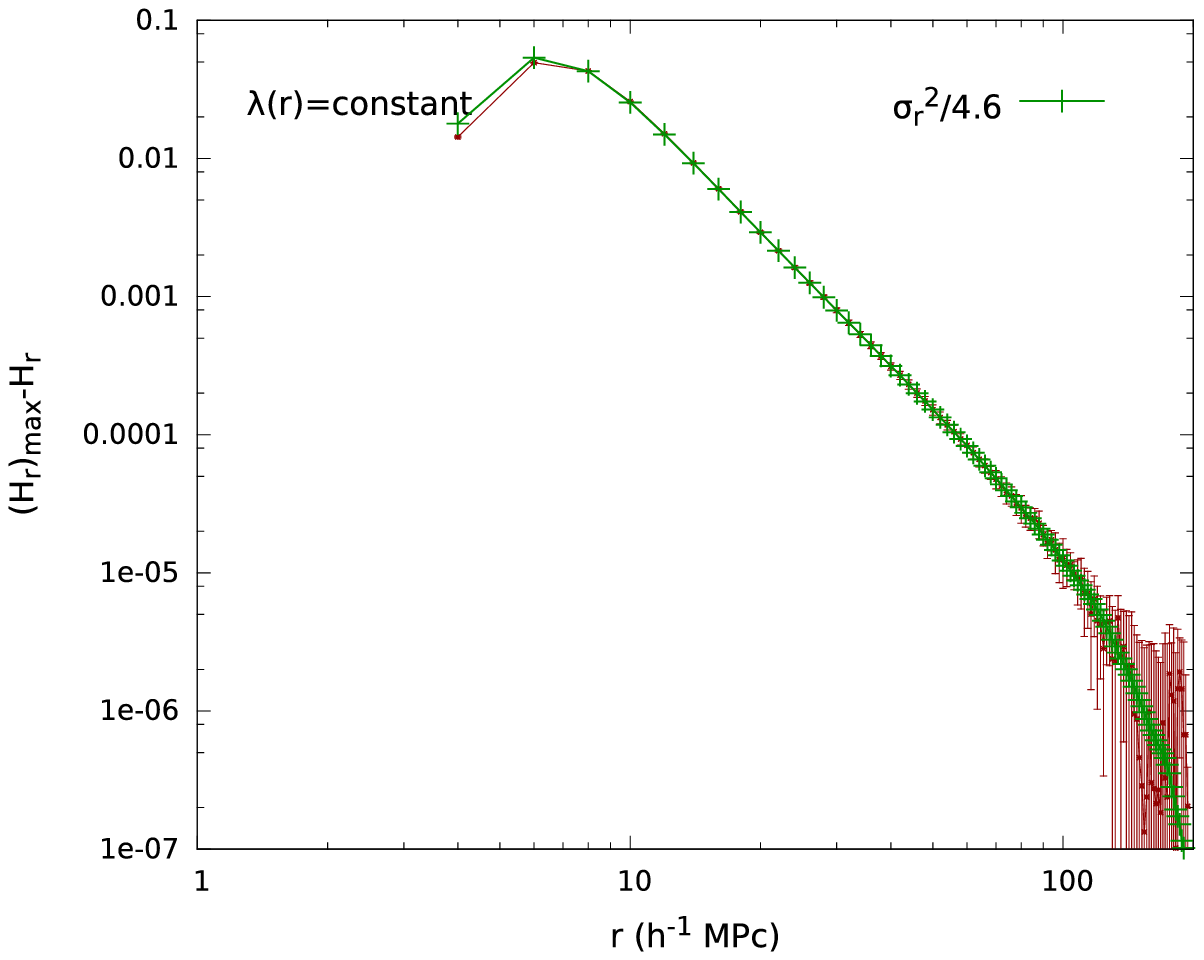}}}%
\resizebox{7cm}{!}{\rotatebox{0}{\includegraphics{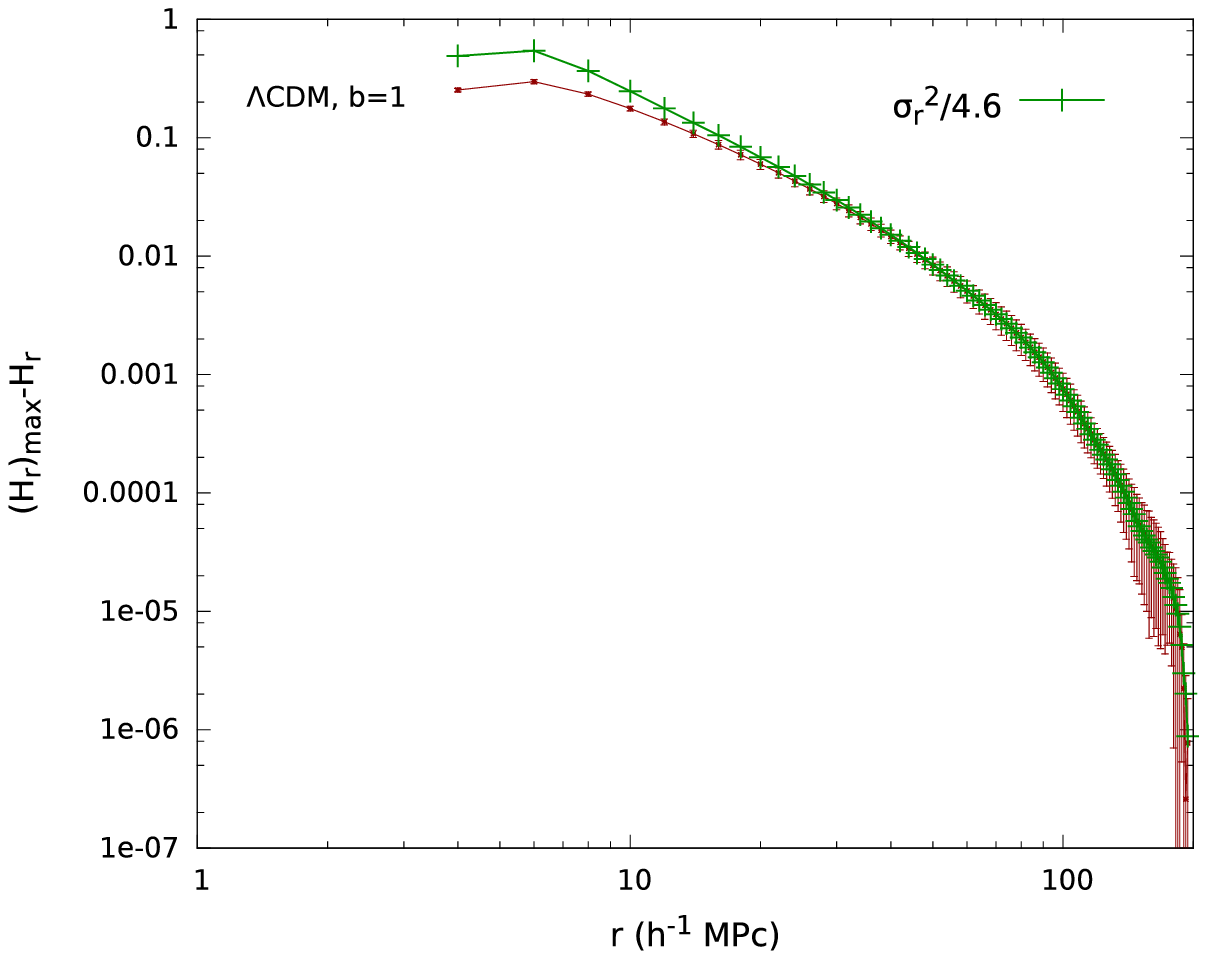}}}\\
\resizebox{7cm}{!}{\rotatebox{0}{\includegraphics{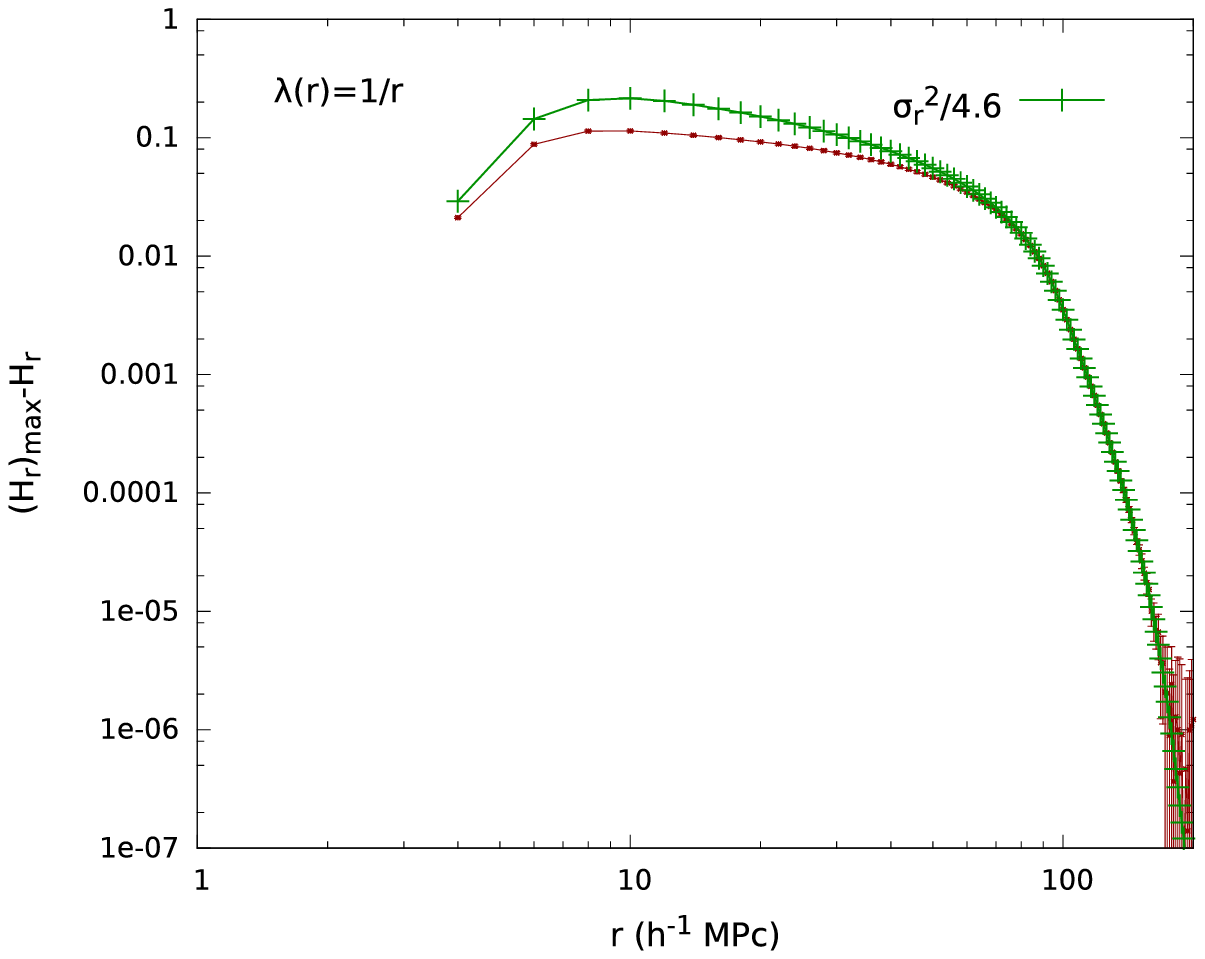}}}%
\resizebox{7cm}{!}{\rotatebox{0}{\includegraphics{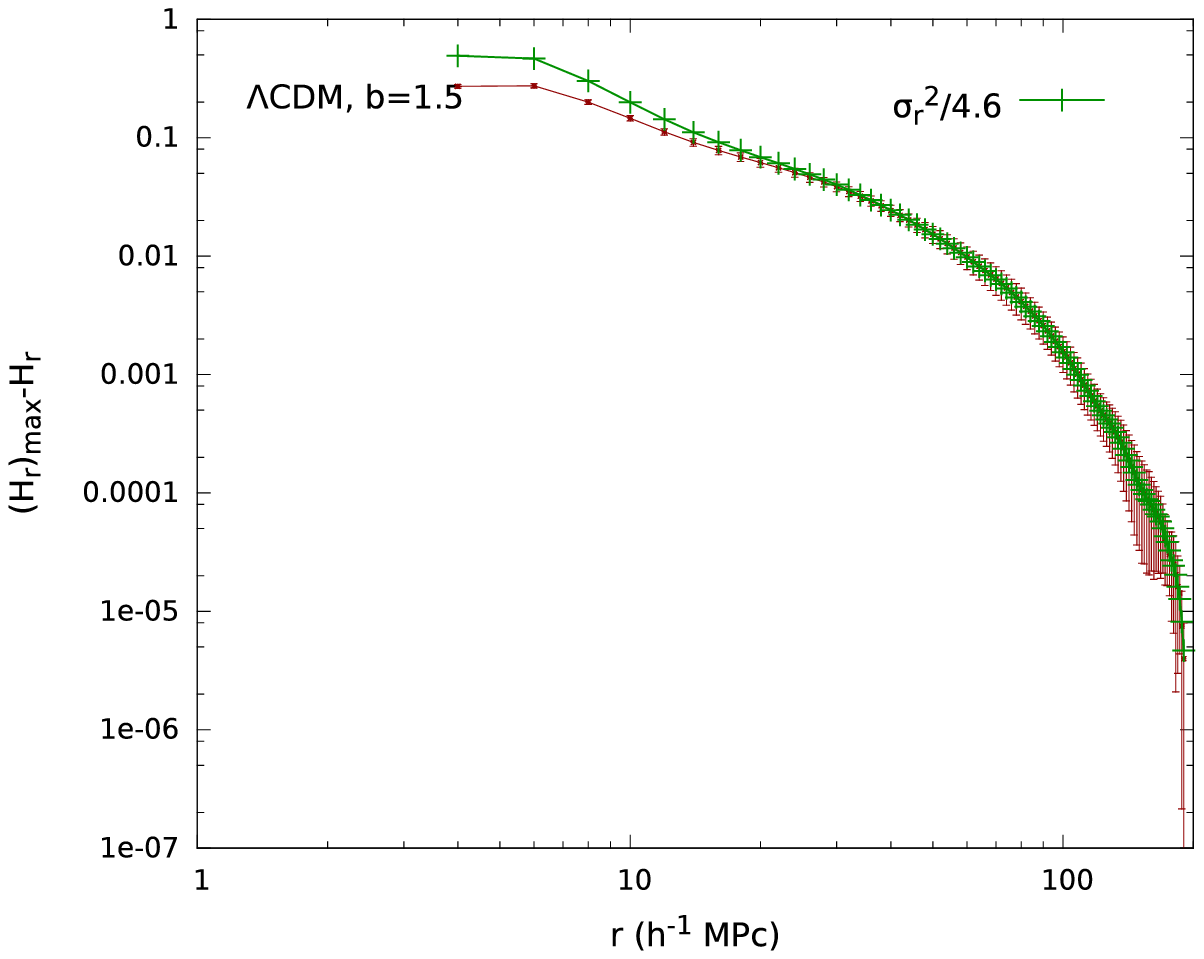}}}\\
\resizebox{7cm}{!}{\rotatebox{0}{\includegraphics{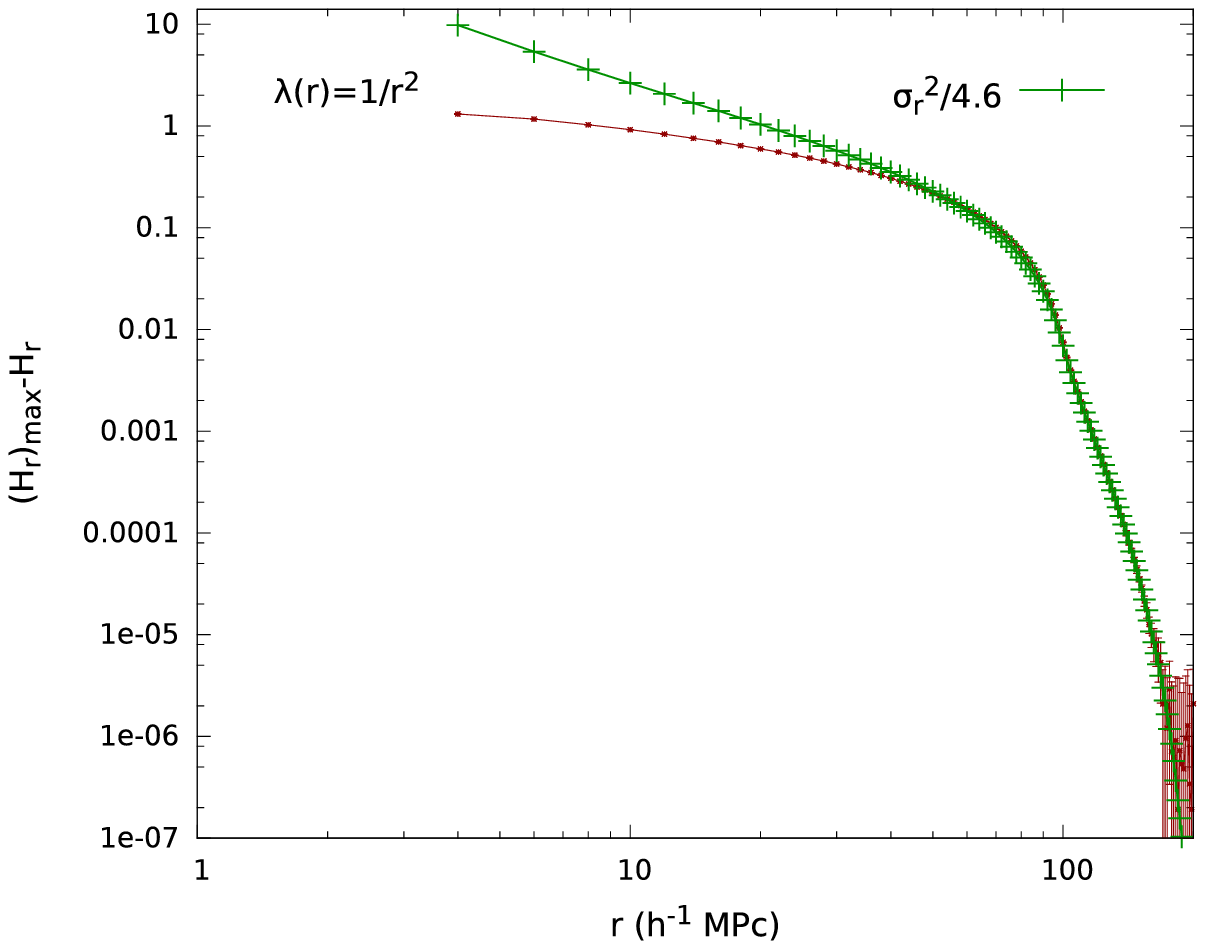}}}%
\resizebox{7cm}{!}{\rotatebox{0}{\includegraphics{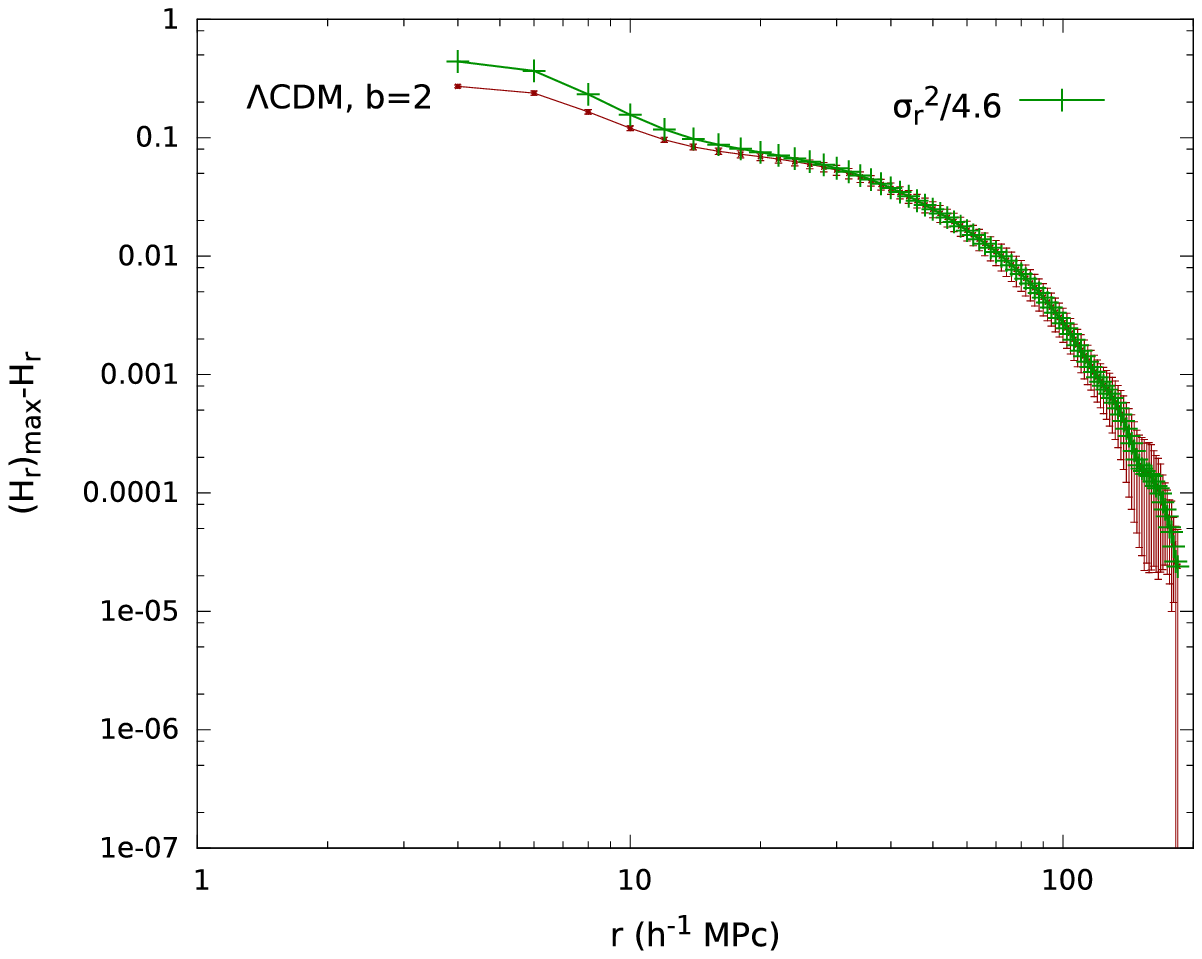}}}\\
\caption{This shows $(H_{r})_{max}-H_{r}$ and
  $\frac{\sigma_{r}^{2}}{4.6}$ as a function of $r$ for different
  distributions as indicated in each panel. In each case we have shown
  only the $1-\sigma$ errorbars for $(H_{r})_{max}-H_{r}$ for
  clarity.}
  \label{fig:result2}
\end{figure*}

\section{DATA}
We use a set of Monte Carlo simulations and N-body simulations to
verify the relation between the information entropy and mass variance.
We use the same data sets to investigate some possible applications of
the information entropy-mass variance relation in cosmology.

\subsection{Monte Carlo simulations of homogeneous and inhomogeneous distributions}

 We generate a set of three dimensional distributions of some
 homogeneous and inhomogeneous distributions using Monte Carlo
 simulations. We consider a set of radial density distributions
 $\rho(r) = K \, \lambda(r)$ where $K$ is a normalization
 constant. The nature of the distribution is governed by the function
 $\lambda(r)$ and we consider $3$ different forms for $\lambda(r)$:
 (i) $\lambda(r) = 1$, (ii) $\lambda(r) = \frac{1}{r}$ and (iii)
 $\lambda(r) = \frac{1}{r^{2}}$. The distribution (i) is a homogeneous
 and isotropic Poisson point process with same density everywhere
 whereas the distributions (ii) and (iii) are inhomogeneous Poisson
 point process. We employ a Monte Carlo dartboard technique to
 simulate these distributions. The detail of the method can be found
 in \citet{pandey13}. Each of the distributions is simulated with
 $N=10^{5}$ points distributed in a spherical region of radius $R=200
 \hmpc$. We generate $10$ such realizations for each of the above
 distributions.

\subsection{N-body simulations}

We use data from a set of N-body simulations of the $\Lambda$CDM model
carried out using a Particle-Mesh (PM) N-body code
\citep{pandey13}. The simulations were run using $256^3$ particles on
a $512^3$ mesh. The simulations cover a comoving volume of $[921.6
  h^{-1} {\rm Mpc}]^3$. We used a $\Lambda$CDM power spectrum with
spectral index $n_s=0.96$ and normalization $\sigma_8=0.812$
(\citealt{komatsu}) with cosmological parameters $\Omega_{m0}=0.27$,
$\Omega_{\Lambda0}=0.73$ and $h=0.71$. The simulations were run for
three different realizations of the initial density fluctuations. We
obtain three different realizations of the dark matter distribution at
$z=0$ from these simulations. In the current paradigm of structure
formation galaxies form at the location of the peaks of the density
field. We use a biasing scheme \citep{cole} where a sharp cutoff is
applied to the smoothed density field allowing the galaxies to form
only in regions where the overdensity exceeds a certain
threshold. Consequently the resulting distributions become biased
relative to the dark matter distributions. We determine the linear
bias parameter $b$ of each simulated biased sample using the ratio,
\begin{equation}
b=\sqrt{ \frac {\xi_g(r)}{\xi_{m}(r)}}
\end{equation}
where $\xi_g(r)$ and $\xi_{m}(r)$ are the two-point correlation
functions for the biased distribution and dark matter distribution
respectively. We generate biased distributions for three different
values of the linear bias $b=1.5$, $b=2$ and $b=2.5$. For each of the
biased and unbiased distributions we consider three non-overlapping
spherical regions of radius $R=200 \, h^{-1} {\rm Mpc}$. We randomly
extract $N=10^5$ particles from each of these spherical regions. This
provides us total nine samples for each bias values.

\begin{figure*}
\resizebox{7cm}{!}{\rotatebox{0}{\includegraphics{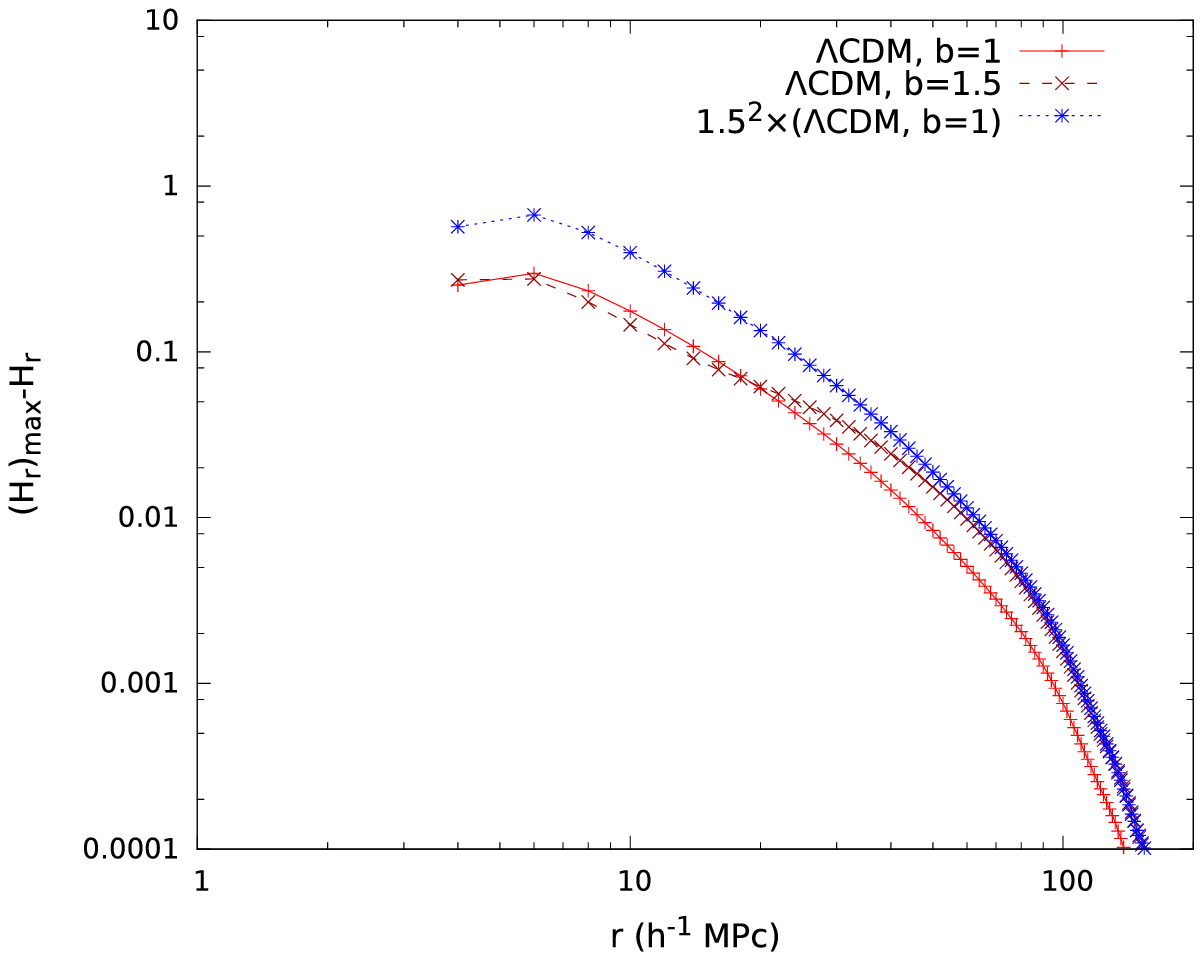}}}%
\resizebox{7cm}{!}{\rotatebox{0}{\includegraphics{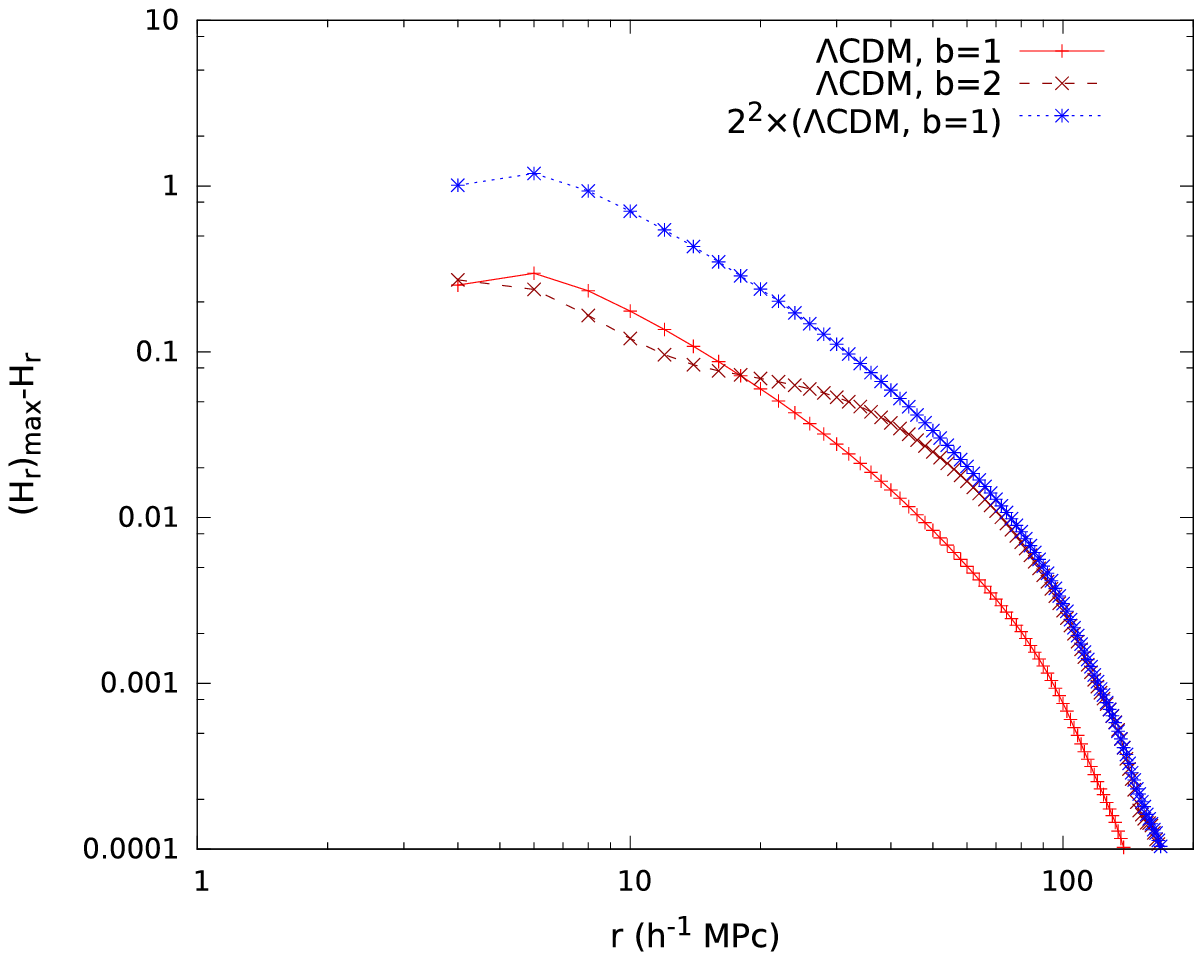}}}\\
\resizebox{7cm}{!}{\rotatebox{0}{\includegraphics{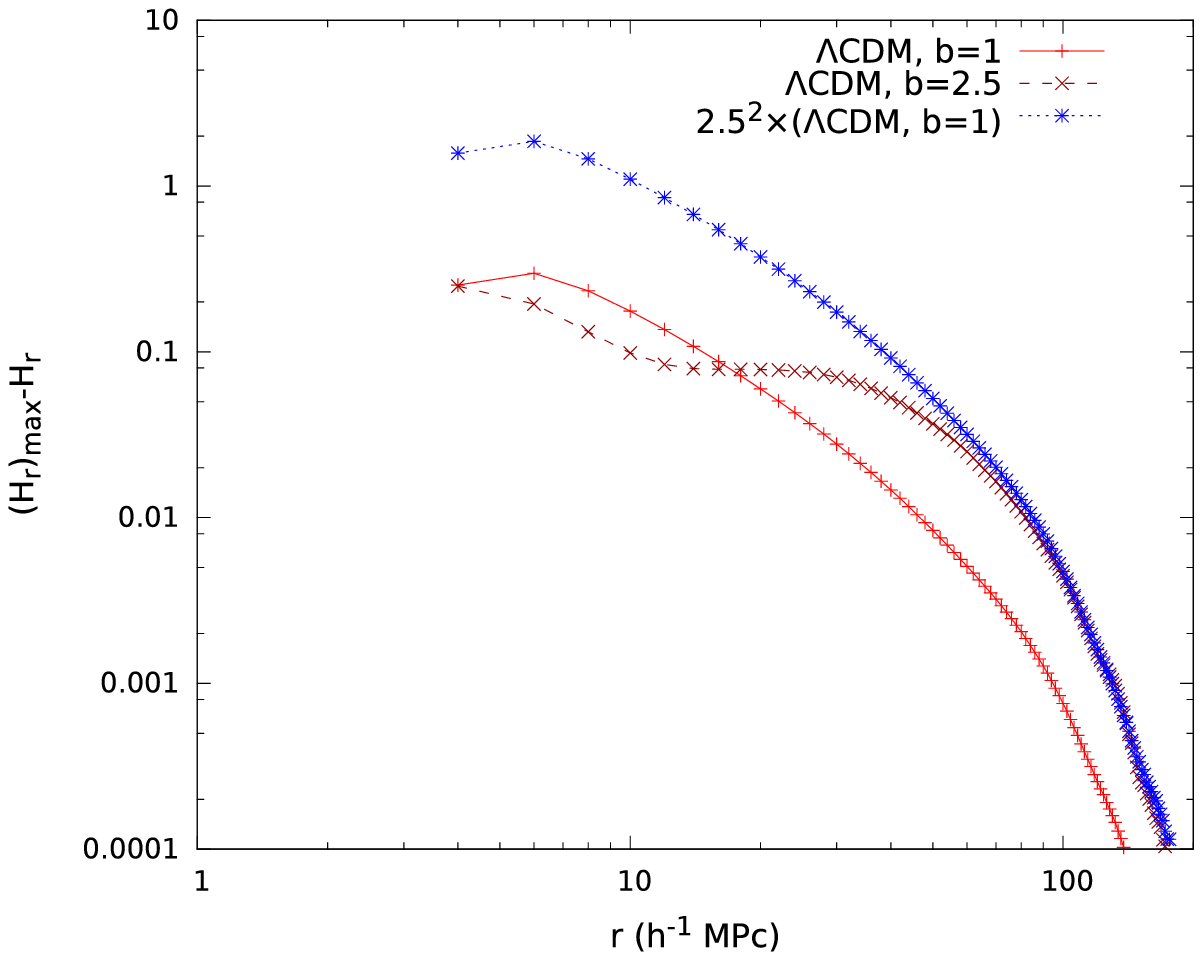}}}%
\resizebox{7cm}{!}{\rotatebox{0}{\includegraphics{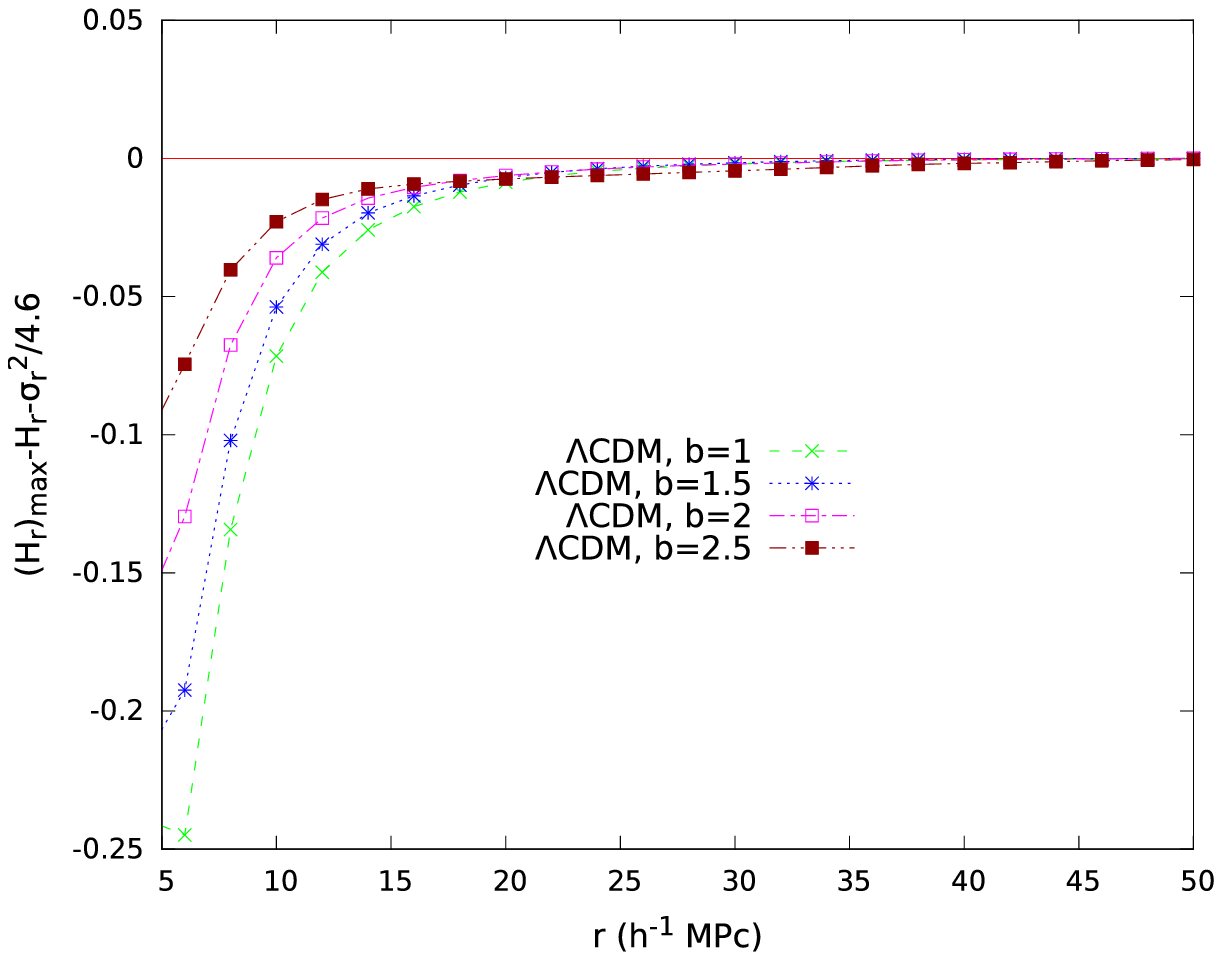}}}\\
\caption{The top left, top right and bottom left panels show that
  scaling the $(H_{r})_{max}-H_{r}$ values for the unbiased
  $\Lambda$CDM model with $b^{2}$ exactly reproduce the
  $(H_{r})_{max}-H_{r}$ values for the different biased $\Lambda$CDM
  model on large scales. This shows that on large scales the ratio of
  $(H_{r})_{max}-H_{r}$ values of the galaxy distributions and dark
  matter distributions from the N-body simulations of the $\Lambda$CDM
  model can be used to determine the linear bias parameter. The bottom
  right panel shows the
  $(H_{r})_{max}-H_{r}-\frac{\sigma_{r}^{2}}{4.6}$ as a function of
  length scales for the unbiased and biased distributions from the
  N-body simulations of the $\Lambda$CDM model. The values of $H_{r}$,
  $(H_{r})_{max}$ and $\sigma_{r}^{2}$ used here are the average
  values estimated from the $9$ samples for each distributions.}
  \label{fig:result3}
\end{figure*}

\section{Results and Conclusions}

We first investigate the relationship between the information entropy
and the mass variance for all the distributions described in section
4. We find that in all the distributions though entropy and $\log
\sigma_{r}$ show some correlations it is difficult to find any general
relation between them. But when we study the relation between the
deviation of entropy from its maximum value i.e. $(H_{r})_{max}-H_{r}$
and the mass variance $\sigma_{r}^2$ for the same distributions,
interestingly we find that they are very tightly correlated
(\autoref{fig:result1}). In \autoref{fig:result1} we find that for a
wide range in their values, the relation between these two quantities
in all these distributions can be described by a straight line of the
form $(H_{r})_{max}-H_{r}=\frac{\sigma_{r}^2}{a}$ where $a$ is a
constant to be determined. We determine the value of $a$ for each of
these distributions by fitting the data with this straight line. We
find that the best fit value of $a=4.65\pm0.03$ is the same for all
the distributions when the data is fitted over $(H_{r})_{max}-H_{r}$
in the range $10^{-5}-10^{-1}$. Some deviations from this relation are
noticed beyond this range. It is interesting to note that in this
regime the relation is exactly described by the relation given in
\autoref{eq:shannvar}. The relation is found to be independent of the
nature of the distribution as predicted by the
\autoref{eq:shannvar}. Further we carry out analysis with different
sampling rates to find that the relation also does not depend on the
number density of the distributions in the appropriate regime as
predicted by the same relation.

In \autoref{fig:result2} we show $(H_{r})_{max}-H_{r}$ and
$\frac{\sigma_{r}^2}{4.6}$ for different distributions as a function
of length scale $r$. In the top left panel we show the results for a
homogeneous and isotropic Poisson point process. We find that for this
distribution the relation holds very well for nearly the entire length
scale range. This is related to the fact that for a homogeneous and
isotropic Poisson distribution the only source of fluctuations are
shot noise which is only important on small scales. We show the
results for the radially inhomogeneous distributions with
$\lambda(r)=\frac{1}{r}$ in the middle left panel and
$\lambda(r)=\frac{1}{r^{2}}$ in the bottom left panel. Here we find
that there are significant deviations from this relation on scales
$r\leq 40 \hmpc$. Interestingly the differences decrease with
increasing length scales and the results are in excellent agreement
with the relation for these distributions beyond the length scales of
$r>50 \hmpc$. In the top right, middle right and bottom right panels
of \autoref{fig:result2} we show the results for the $\Lambda$CDM
model with different linear bias values as indicated in each panel. We
find small departures from the relation in each of these distributions
on smaller length scales $r\leq 20 \hmpc$ but the relation holds
astonishingly well on length scales of $r>20 \hmpc$. It may be noted
in different panels of \autoref{fig:result2} that the shape of the
$(H_{r})_{max}-H_{r}$ curves are quite different from each other which
are the characteristics of the respective distributions. But the
relation given in \autoref{eq:shannvar} holds quite well irrespective
of the nature of the distributions. The deviations of the results from
\autoref{eq:shannvar} in all these distributions originate from the
presence of larger fluctuations on those length scales. This retains
the non-vanishing higher order terms in \autoref{eq:shannon3} giving
rise to those differences. But the higher order terms become
negligible in the small fluctuation regime where \autoref{eq:shannvar}
becomes exact. Therefore deviation from the \autoref{eq:shannon3}
provides the degree of non-linearities present and the length scales
where the non-linearities become important in a distribution.

We now investigate if the information entropy-mass variance relation
can be used to determine the linear bias and characterize the
non-Gaussianities. In the top left panel of \autoref{fig:result3} we
show the $(H_{r})_{max}-H_{r}$ values as a function of length scales
for the unbiased ($b=1$) $\Lambda$CDM simulation and its biased
counterpart with the bias value $b=1.5$.  We see that when the
$(H_{r})_{max}-H_{r}$ values for the unbiased distributions are scaled
by a factor of $1.5^{2}$, it exactly reproduces the
$(H_{r})_{max}-H_{r}$ values for the biased distributions on scales
$r>30 \hmpc$. The top right and bottom left panels of
\autoref{fig:result3} similarly show that on scales $r>30 \hmpc$ the
$(H_{r})_{max}-H_{r}$ values for the biased distributions with $b=2$
and $b=2.5$ are simply $2^{2}$ and $2.5^{2}$ times the
$(H_{r})_{max}-H_{r}$ values of the unbiased distributions on those
scales. However it can be clearly seen in all these panels that this
simple scaling does not work on smaller scales. The assumption of the
scale independent linear bias does not hold on small scales where the
non-linearities play an important role. The particles are distributed
in diverse environments in an unbiased distribution whereas they are
preferentially selected from the density peaks in a biased
distribution. In a biased distribution the measuring spheres centered
on the particles would encompass regions with similar densities at
smaller radii. But the measuring spheres would encompass varying
degrees of empty regions beyond the characteristic scales of the
density peaks leading to non-uniformity in the measurements with
increasing radii. On the other hand, in an unbiased distribution, the
measuring spheres would pick up regions of diverse densities at
smaller radii reflecting a less uniform behaviour on those scales. The
unbiased distribution would be more uniform at larger radii when the
measuring spheres would encompass statistically similar number of
sites from different environments. This explains why the biased
distributions appear to be more uniform on smaller scales and less
uniform on larger scales as compared to the unbiased
distributions. These characteristic behaviours of the biased
distributions give rise to the observed differences from the
\autoref{eq:biasval} on smaller scales. Despite these differences it
is clear that one can use the ratios of $(H_{r})_{max}-H_{r}$ values
of different distributions on large scales to determine their relative
bias parameters. The method can be also used to determine the linear
bias for galaxies with different physical properties. In future we
plan to study the luminosity-bias relation for the galaxies in the
SDSS using this method.

In the bottom right panel of \autoref{fig:result3} we show
$(H_{r})_{max}-H_{r}-\frac{\sigma_{r}^{2}}{4.6}$ as a function of
length scales for the biased and the unbiased distributions considered
here. We find that $(H_{r})_{max}-H_{r}-\frac{\sigma_{r}^{2}}{4.6}<0$
for all the distributions upto a length scales of $\sim 40 \hmpc$
suggesting that all of them are non-Gaussian. We see that it is more
negative in the unbiased distributions as compared to the biased
distributions. It may be also noted that the measure becomes less
negative with increasing bias. This behaviour is possibly related to
the fact that a biased distribution becomes more uniform on small
scales with increasing bias. However as this measure is a combination
of different odd and event moments with alternating signs, it is
difficult in general to absolutely compare the degree of
non-Gaussianity present in these distributions.

In this work we present a relation between the information entropy and
the mass variance and show that on large scales the relation can be
used to determine the linear bias from galaxy surveys. The relation
may be also employed to constrain the growth rate of density
fluctuations and time evolution of linear bias on large scales. On
small scales one can use the relation to characterize the
non-Gaussianities present in the galaxy distribution. Finally we note
that the present analysis suggests that the information entropy can
serve as an important tool for the study of large scale structures in
the Universe.

\section{Acknowledgement}
I sincerely thank an anonymous referee for constructive comments and
suggestions which helped me to significantly improve the draft. The
author would like to acknowledge IUCAA, Pune and CTS, IIT Kharagpur
for the use of its facilities for the present work. The author would
also like to thank Shreekantha Sil for useful comments and
discussions.

\bsp	
\label{lastpage}
\end{document}